\documentclass[runningheads]{llncs}
\usepackage{graphicx}
\usepackage{bm}
\usepackage{amssymb,amsmath}
\usepackage{cite}
\usepackage{color}
\usepackage{soul}

\begin{document}
\title{Physics-Informed Echo State Networks for Chaotic Systems Forecasting\thanks{The authors acknowledge the support of the Technical University of Munich - Institute for Advanced Study, funded by the German Excellence Initiative and the European Union Seventh Framework Programme under grant agreement no. 291763. L.M. also  acknowledges the Royal Academy of Engineering Research Fellowship Scheme.}}
%
%
\author{N.A.K. Doan\inst{1,2} \and
W. Polifke\inst{1} \and
L. Magri\inst{2,3}
}
\authorrunning{N.A.K. Doan et al.}
%
\institute{Department of Mechanical Engineering, Technical University of Munich, Germany \and
Institute for Advanced Study, Technical University of Munich, Germany\\
\and
Department of Engineering, University of Cambridge, United Kingdom}
\maketitle              
\begin{abstract}

We propose a physics-informed Echo State Network (ESN) to predict the evolution of chaotic systems. Compared to conventional ESNs, the physics-informed ESNs are trained to solve supervised learning tasks while ensuring that their predictions do not violate physical laws. This is achieved by introducing an additional loss function during the training of the ESNs, which penalizes non-physical predictions without the need of any additional training data. This approach is demonstrated on a chaotic Lorenz system, where the physics-informed ESNs improve the predictability horizon by about two Lyapunov times as compared to conventional ESNs.
The proposed framework shows the potential of using machine learning combined with prior physical knowledge to improve the time-accurate prediction of chaotic dynamical systems.


\keywords{Echo State Networks \and Physics-Informed Neural Networks \and Chaotic Dynamical Systems.}
\end{abstract}

\section{Introduction}

Over the past few years, there has been a rapid increase in the development of machine learning techniques, which have been applied with success to various disciplines, from image or speech recognition \cite{Krizhevsky2012,Hinton2012} to playing Go \cite{Silver2016}. However, the application of such methods to the study and forecasting of physical systems has only been recently explored including some applications in the field of fluid dynamics \cite{Raissi2019a,Ling2016,Jaensch2017,Wu2018}.
One of the major challenges for using  machine learning algorithms for the study of complex physical systems is the prohibitive cost of data acquisition and generation for training \cite{Duraisamy2018,Raissi2019}. However, in complex physical systems, there exists a large amount of prior knowledge, which can be exploited to improve  existing machine learning approaches. These hybrid approaches, called \emph{physics-informed machine learning}, have been applied with some success to flow-structure interaction problems \cite{Raissi2019}, turbulence modelling \cite{Ling2016} and the solution of partial differential equations (PDEs) \cite{Raissi2019}.

In this study, we propose an approach to combine physical knowledge with a machine learning algorithm to time-accurately forecast the evolution of a chaotic dynamical system. The machine learning tools we use are based on reservoir computing \cite{Lukosevicius2009}, in particular,  Echo State Networks (ESNs). ESNs have shown to predict nonlinear and chaotic dynamics more accurately and for a longer time horizon than other deep learning algorithms \cite{Lukosevicius2009}. ESNs have also recently been used to predict the evolution of spatiotemporal chaotic systems \cite{Pathak2018a,Pathak2018}. In the present study, ESNs are augmented by physical constraints to accurately forecast the evolution of a prototypical chaotic system, i.e., the Lorenz system \cite{Lorenz1963}.

Sections~\ref{sec:method} details the method used for the training and for forecasting the dynamical systems, both with conventional ESNs and the newly proposed physics-informed ESNs. Results are presented in section \ref{sec:results} and final comments are summarized in section \ref{sec:conclusion}.

\section{Methodology}
\label{sec:method}
The Echo State Network (ESN) approach presented in \cite{Lukosevicius2012} is used here. Given a training input signal $\bm{u}(n)$ of dimension $N_u$ and a desired known target output signal $\bm{y}(n)$ of dimension $N_y$, the ESN has to learn a model with output $\widehat{\bm{y}}(n)$ matching $\bm{y}(n)$. $n=1, ..., N_t$ is the discrete time and $N_t$ is the number of data points in the training dataset covering a discrete time from $0$ until time $T=(N_t-1)\Delta t$. In the particular case studied here, where the forecasting of a dynamical system is of interest, the desired output signal is equal to the input signal at the next time step, i.e., $\bm{y}(n) = \bm{u}(n+1)$.

The ESN is composed of an artificial high dimensional dynamical system, called a reservoir, whose states at time $n$ are represented by a vector, $\bm{x}(n)$, of dimension $N_x$, representing the reservoir neuron activations. This reservoir is coupled to the input signal, $\bm{u}$, via an input-to-reservoir matrix, $\bm{W}_{in}$ of dimension $N_x \times N_u$. The output of the reservoir, $\widehat{\bm{y}}$, is deduced from the states via the reservoir-to-output matrix, $\bm{W}_{out}$ of dimension $N_y \times N_x$, as a linear combination of the reservoir states:
\begin{equation}
    \widehat{\bm{y}}=\bm{W}_{out} \bm{x}
\end{equation}
In this work, a non-leaky reservoir is used, where the state of the reservoir evolves according to:
\begin{equation}
\bm{x}(n+1) = \tanh \left( \bm{W}_{in} \bm{u}(n) + \bm{W} \bm{x}(n) \right)
\end{equation}
where $\bm{W}$ is the recurrent weight matrix of dimension $N_x \times N_x$ and the (element-wise) $\tanh$ function is used as an activation function for the reservoir neurons.

\begin{figure}[!ht]
	\centering
	\includegraphics[width=0.65\textwidth]{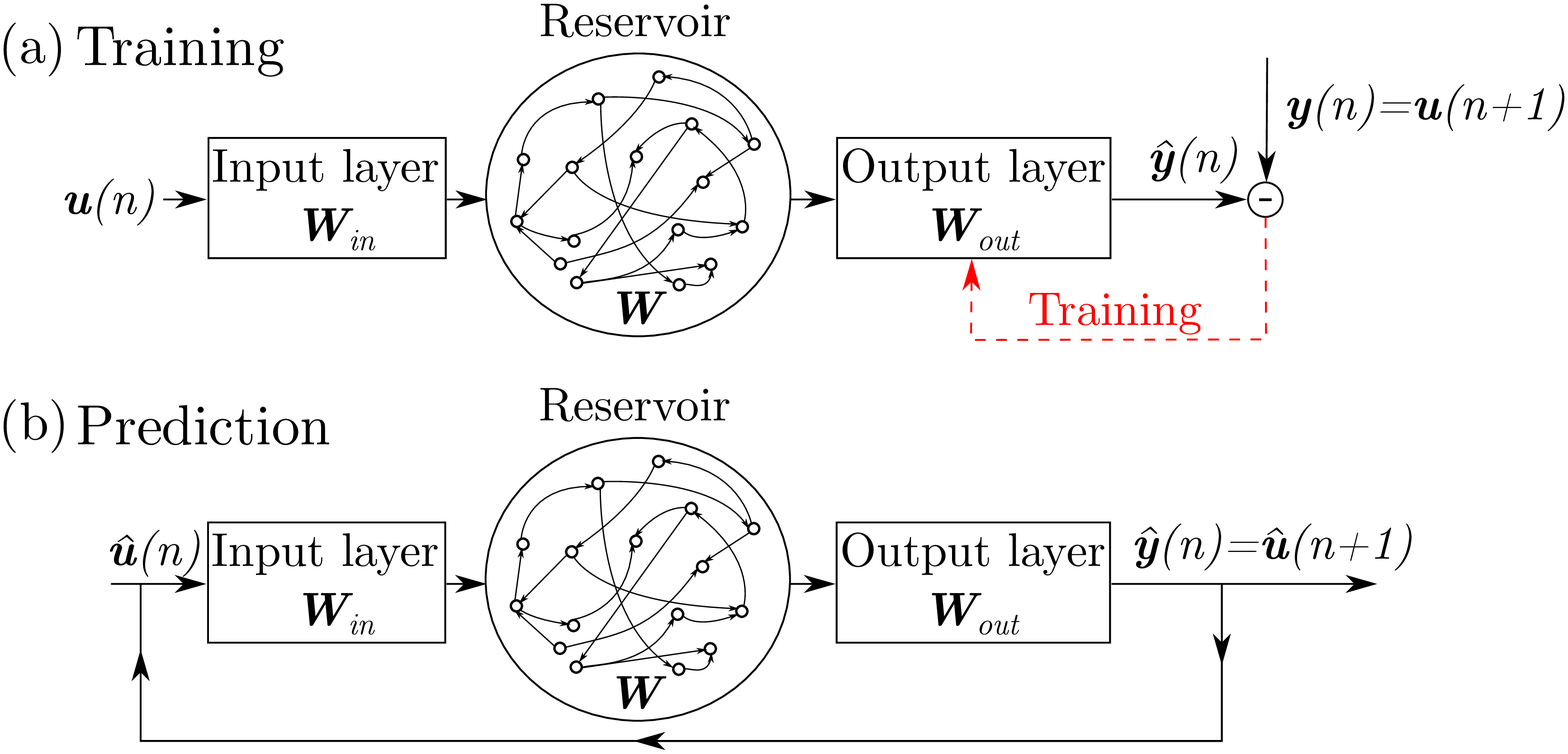}
	\caption{Schematic of the  ESN during (a) training and (b) future prediction. The physical constraints are imposed during the training phase (a). }
	\label{fig:ESN_schema}
\end{figure}

In the conventional ESN approach, illustrated in Fig. \ref{fig:ESN_schema}a, the input and recurrent matrices, $\bm{W}_{in}$ and $\bm{W}$, are randomly initialized once and are not further trained. These are typically sparse matrices constructed so that the reservoir verifies the Echo State Property \cite{Jaeger2004}. Only the output matrix, $\bm{W}_{out}$, is then  trained to minimize the mean-square-error, $E_d$, between the ESN predictions and the data defined as:
\begin{equation}
E_{d} = \frac{1}{N_y} \sum_{i=1}^{N_y} \frac{1}{N_t} \sum_{n=1}^{N_t} (\widehat{y}_i (n) - y_i (n) )^2 \label{eq:err_data}
\end{equation}
The subscript $d$ is used to indicate the error based on the available data. 

In the present implementation, following \cite{Pathak2018a}, $\bm{W}_{in}$ is generated such that each row of the matrix has only one randomly chosen nonzero element, which is independently taken from a uniform distribution in the interval $[-\sigma_{in}, \sigma_{in}]$. $\bm{W}$ is constructed to have an average connectivity $\langle d \rangle$ and the non-zero elements are taken from a uniform distribution over the interval $[-1,1]$. All the coefficients of $\bm{W}$ are then multiplied by a constant coefficient for the largest absolute eigenvalue of $\bm{W}$ to be equal to a value $\Lambda$.

After training, to obtain predictions from the ESN for future time $t>T$, the output of the ESN is looped back as the input of the ESN and it then evolves autonomously as represented in Fig. \ref{fig:ESN_schema}b.

\subsection{Training}
As discussed earlier, the training of the ESN consists of the optimization of $\bm{W}_{out}$. As the outputs of the ESN, $\widehat{\bm{y}}$, are a linear combination of the states, $\bm{x}$, $\bm{W}_{out}$ can be obtained by using a simple Ridge regression:

\begin{equation}
\bm{W}_{out} = \bm{Y} \bm{X}^T \left( \bm{X} \bm{X}^T + \gamma \bm{I}  \right)^{-1} \label{eq:ridge}
\end{equation}
where $\bm{Y}$ and $\bm{X}$ are respectively the column-concatenation of the various time instants of the output data, $\bm{y}$, and associated ESN states $\bm{x}$. $\gamma$ is the Tikhonov regularization factor, which helps avoid overfitting. Explicitly, the optimization in Eq. (\ref{eq:ridge}) reads:

\begin{equation}
\bm{W}_{out} = \underset{\bm{W}_{out}}{\text{argmin}} \frac{1}{N_y} \sum_{i=1}^{N_y} \left( \sum_{n=1}^{N_t} (\widehat{y}_i(n) - y_i(n) )^2 + \gamma || \bm{w}_{out,i} ||^2 \right)
\end{equation}
where $\bm{w}_{out,i}$ denotes the $i$-th row of $\bm{W}_{out}$. This optimization problem penalizes large values of $\bm{W}_{out}$, generally improves the feedback stability and  avoids overfitting \cite{Lukosevicius2012}.

In this work, following the approach of \cite{Raissi2019} for artificial deep feedforward neural networks, we propose an alternative approach to training $\bm{W}_{out}$, which combines the data available with prior physical knowledge of the system under investigation. Let us first assume that the dynamical system is governed by the following nonlinear differential equation:
\begin{equation}
\mathcal{F}(\bm{y}) \equiv \partial_t \bm{y} + \mathcal{N} (\bm{y}) = 0
\label{eq:dynamic}
\end{equation}
where $\mathcal{F}$ is a general non-linear operator, $\partial_t$ is the time derivative and $\mathcal{N}$ is a nonlinear differential operator. Eq. (\ref{eq:dynamic}) represents a formal equation describing the dynamics of a generic nonlinear system. The training phase can be reframed to make use of our knowledge of $\mathcal{F}$ by minimising the mean squared error, $E_d$, and a physical error, $E_p$, based on $\mathcal{F}$:
\begin{equation}
E_{tot} = E_d + E_p, \text{~where~} E_p = \frac{1}{N_y} \sum_{i=1}^{N_y}  \frac{1}{N_p} \sum_{p=1}^{N_p} | \mathcal{F}(\widehat{y_i}(n_p))|^2
\label{eq:Etot}
\end{equation}
Here, the set $\lbrace \widehat{\bm{y}} (n_p)  \rbrace_{p=1}^{N_p}$ denotes ``collocation points" for $\mathcal{F}$, which are defined as a prediction horizon of $N_p$ datapoints obtained from the ESN covering the time period $(T+\Delta t) \leq t \leq (T+N_p \Delta t)$. Compared to the conventional approach where the regularization of $\bm{W}_{out}$ is based on avoiding extreme values of $\bm{W}_{out}$, our proposed method regularizes $\bm{W}_{out}$ by using our prior physical knowledge. Equation (\ref{eq:Etot}), which is a key equation, shows how to constrain our prior physical knowledge in the loss function.
Therefore, this procedure ensures that the ESN becomes predictive because of  data training and the ensuing prediction is consistent with the physics.
It is motivated by the fact that in many complex physical systems, the cost of data acquisition is prohibitive and thus, there are many instances where only a small amount of data is available for the training of neural networks. In this context, most existing machine learning approaches lack robustness: Our approach better leverages on the information content of the data that the machine learning algorithm uses. 
Our physics-informed framework is straightforward to implement because it only requires the evaluation of the residual, but it does not require the computation of the exact solution.

\section{Results}
\label{sec:results}
The approach described in section \ref{sec:method} is applied for forecasting the chaotic evolution of the chaotic Lorenz system, which is described by the following equations \cite{Lorenz1963}:
\begin{equation}
    \frac{du_1}{dt} = \sigma (u_2-u_1), \hspace*{11pt}
    \frac{du_2}{dt} = u_1 (\rho-u_3)-u_2, \hspace*{11pt}
    \frac{du_3}{dt} = u_1 u_2-\beta u_3 \label{eq:Lorenz}
\end{equation}
where $\rho=28$, $\sigma = 10$ and $\beta=8/3$. These are the standard values of the Lorenz system that spawn a chaotic solution \cite{Lorenz1963}. The size of the training dataset is $N_t=1000$ and the timestep between two time instants is $\Delta t = 0.01$.

The parameters of the reservoir both for the conventional and physics-informed ESNs are taken to be: $\sigma_{in} = 0.15$, $\Lambda = 0.4$ and $\langle d \rangle = 3$. In the case of the conventional ESN, the value of $\gamma = 0.0001$ is used for the Tikhonov regularization. These values of the hyperparameters are taken from previous studies \cite{Pathak2018,Pathak2018a}.

For the physics-informed ESN, a prediction horizon of $N_p=1000$ points is used and the physical error is estimated by discretizing Eq. (\ref{eq:Lorenz}) using an explicit Euler time-integration scheme. The choice of $N_p=1000$ gives equal importance to the error based on the data and the error based on the physical constraints. The optimization of $\bm{W}_{out}$ is performed using the L-BFGS-B algorithm with the $\bm{W}_{out}$ obtained by Ridge regression (Eq. (\ref{eq:ridge})) as the initial guess.

\begin{figure}[!ht]
	\centering
	\includegraphics[width=0.8\textwidth]{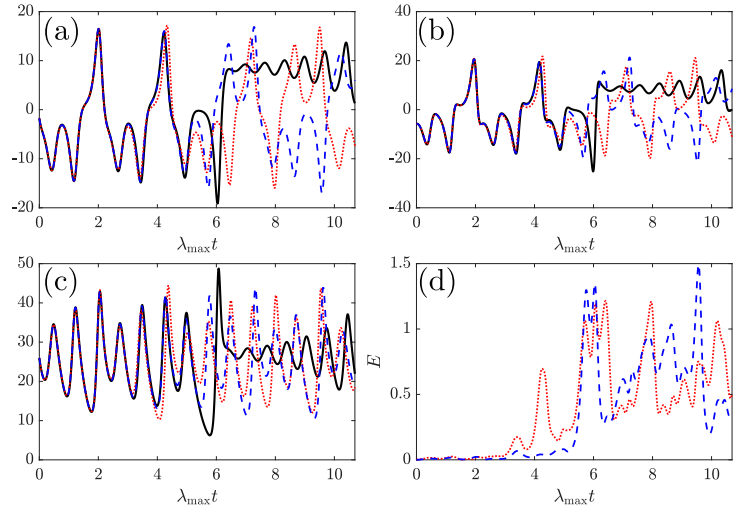}
	\caption{Prediction of the Lorenz system (a) $u_1$, (b) $u_2$, (c) $u_3$ and (d) $E$ using the conventional ESN (dotted red lines) and the physics-informed ESN (dashed blue lines). The actual evolution of the Lorenz system is shown using full black lines.}
	\label{fig:Lorenz_200U}
\end{figure}

\begin{figure}[!ht]
	\centering
	\includegraphics[width=0.65\textwidth]{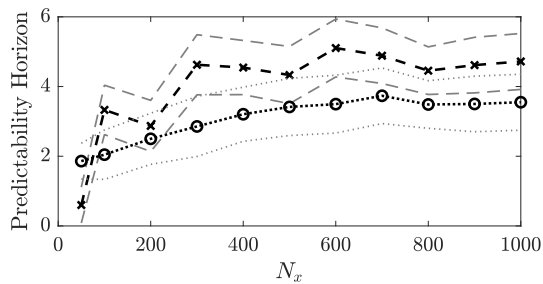}
	\caption{Mean predictability horizon of the conventional ESN (dotted line with circles) and physics-informed ESN (dashed line with crosses) as a function of the reservoir size ($N_x$). The associated gray lines indicate the standard deviation from the mean.}
	\label{fig:valid_time}
\end{figure}

 The predictions for the Lorenz system by conventional and physics-informed ESNs are compared with the actual evolution in Fig. \ref{fig:Lorenz_200U}, where the time is normalized by the largest Lyapunov exponent, $\lambda_{\max}=0.934$, and the reservoir has 200 units. Figure \ref{fig:Lorenz_200U}d shows the evolution of the normalized error, which is defined as
\begin{equation}
E(n) = \frac{|| \bm{u}(n) - \widehat{\bm{u}}(n)||}{\langle || \bm{u} ||^2 \rangle^{1/2}}
\end{equation}
where $\langle \cdot \rangle$ denotes the time average.
The physics-informed ESN shows a remarkable improvement of the time over  which the predictions are accurate. Indeed, the time for the normalized error to exceed 0.2, which is the threshold used here to define the predictability horizon, improves from 4 Lyapunov times to approximately 5.5 with the physics-informed ESN. 
The dependence of the predictability horizon on the reservoir size is estimated as follows (Fig. \ref{fig:valid_time}). First, the trained physics-informed and conventional ESNs are run for an ensemble of 100 different initial conditions. Second, for each run, the predictability horizon is calculated. Third, the mean and standard deviation of the predictability horizon are computed from the ensemble. It is observed that the physics-informed approach provides a marked improvement of the predictability horizon over conventional ESNs and, most significantly, for reservoirs of intermediate sizes. The only exception is for the smallest reservoir ($N_x=50$). 
In principle, it may be conjectured that a conventional ESN may have 
 a similar performance to that of a physics-informed ESN by ad-hoc optimization of the  hyperparameters. However, no efficient methods are available (to date) for  hyperparameter optimization \cite{Lukosevicius2009}. The approach proposed here allows us to improve the performance of the ESN (optimizing $\bm{W}_{out}$) by adding a constraint on the physics, i.e., the governing equations, and not by ad-hoc tuning of the hyperparameters. This suggests that the physics-informed approach is more robust than the conventional approach.

\section{Conclusions and future directions}
\label{sec:conclusion}
We propose an approach for training echo state networks (ESNs), which constrains the knowledge of the physical equations that govern a dynamical system. This physics-informed approach is shown to be more robust than purely data-trained ESNs: The predictability horizon is markedly increased without requiring additional training data.
In ongoing work, 
(i) the impact that the number of collocation points has on the accuracy of the ESNs is thoroughly assessed; 
(ii) the reason why the predictability horizon saturates as the reservoirs become larger is investigated; and  (iii) the physics-informed ESNs are applied to high dimensional fluid dynamics systems. Importantly, the physics-informed framework we propose will be exploited to minimize the data required for training. 
This work opens up new possibilities for the time-accurate prediction of the dynamics of chaotic systems by constraining the underlying physical laws.

%
\bibliographystyle{splncs04}

\end{document}